\documentclass[final]{siamltex}
\usepackage{graphicx}
\usepackage{epstopdf}

\usepackage{amsmath} \interdisplaylinepenalty=2500
\usepackage{amsgen}
\usepackage{amsfonts}
\usepackage{amssymb}
\usepackage{amsbsy}
\usepackage{bm}  

\usepackage{showkeys}

\newcommand{\bgamma}{\boldsymbol{{\gamma}}}
\newcommand{\bGamma}{\boldsymbol{{\Gamma}}}

\newcommand{\bdot}{\boldsymbol{\cdot}}

\def\d{\,{\rm d}}
\def\i{{\rm i}}
\def\e{{\rm e}}

\begin{document}

\title{\bf Theory of Waveform-Diverse Moving-Target Spotlight Synthetic-Aperture Radar }


\author{Margaret Cheney\thanks{Department of Mathematical Sciences, Rensselaer Polytechnic Institute, Troy, NY  12180 USA} \and Brett Borden\thanks{$^2$Physics Department, Naval Postgraduate School, Monterey, CA 93943-5001 USA}}

\maketitle

\begin{abstract}
We develop a theory for waveform-diverse moving-target synthetic-aperture radar, in the case in which a single moving antenna is used for both transmitting and receiving. We assume that the targets (scattering objects) are moving linearly, but we allow an arbitrary, known flight path for the antenna and allow it to transmit a sequence of arbitrary, known waveforms.

A formula for phase space (position and velocity) imaging is developed, and we provide a formula for the point-spread function of the corresponding imaging system.  This point-spread function is expressed in terms of the ordinary radar ambiguity function.  
  
As an example, we show how the theory can be applied to the problem of estimating the errors that arise when target and antenna motion is neglected during the transit time of each pulse.  
\end{abstract}

\section{Introduction}
Synthetic-aperture radar ({\sc sar}) images are formed using a moving antenna which transmits pulses of electromagnetic energy and measures the field scattered by objects (targets) in the environment.  The measurements from different antenna locations are combined --- typically by backprojection --- and the resulting reflectivity map (target image) can display much finer resolution than that available from measurements at a single antenna location. (See, e.g. \cite{CGM,CBbook,CS,El,FL,R,S}.) Standard {\sc sar} systems transmit the same high-range-resolution waveform repeatedly in order to measure the range (distance) from the antenna to various scene components.

Below, we develop the theory for a synthetic-aperture radar that makes use of a variety of different waveforms.  We consider the case when there are multiple targets moving  independently in the scene while the radar data are acquired. We assume that the targets are moving linearly, but we allow the antenna to fly an arbitrary, known, flight path and allow it to transmit a sequence of arbitrary, known waveforms. Our development is restricted to {\em monostatic} radar configurations, in which a single antenna is used for both transmitting and receiving. The problem is formulated in terms of forming an image in phase space, where the independent variables include not only the position of the target but also its vector velocity.

\begin{figure}[ht]
\centering    
\includegraphics[width=3in]{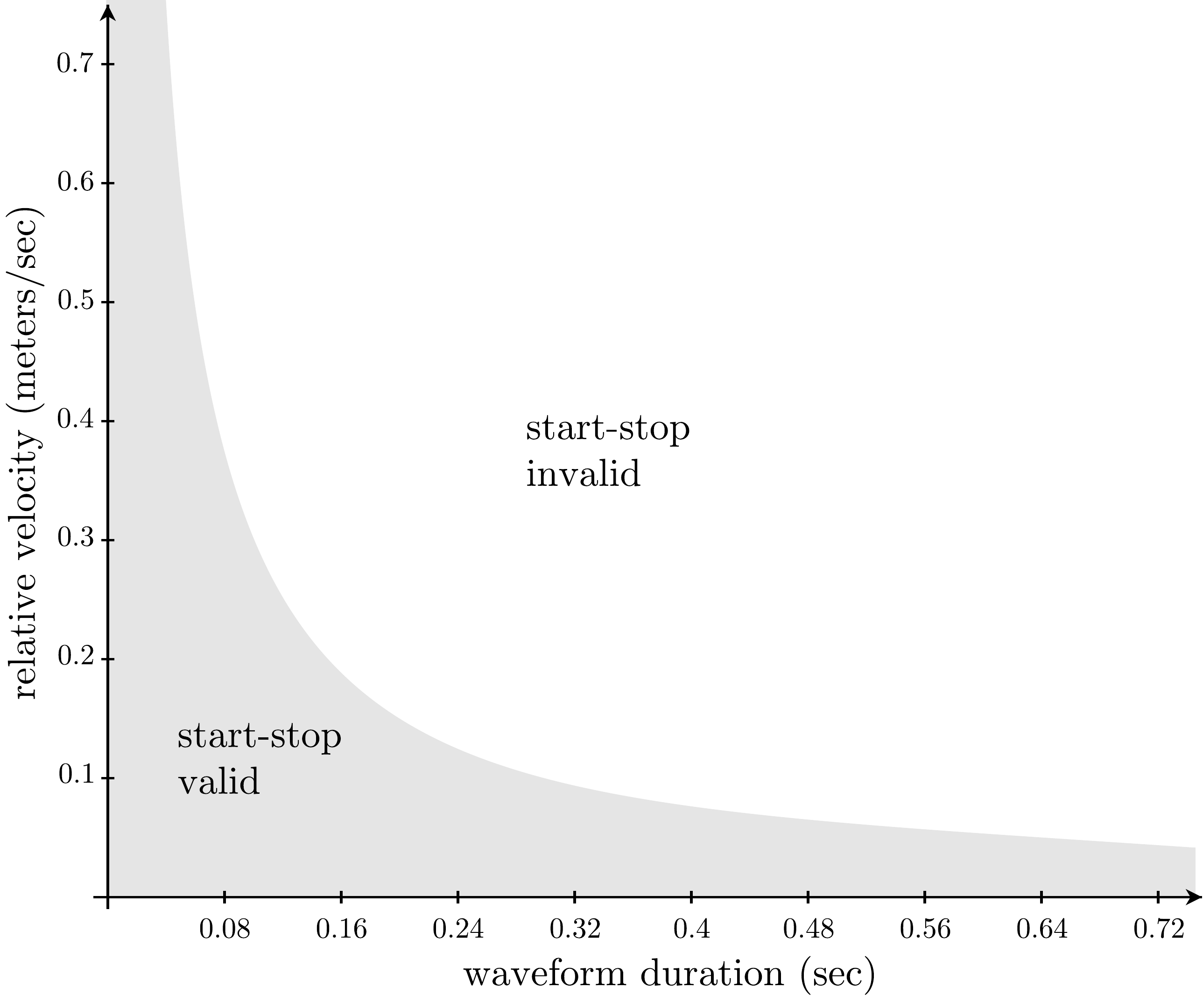} 
\caption{The approximate region of validity for the start-stop approximation for X-band (wavelength 3cm).  The horizontal axis represents the waveform duration, and the vertical axis shows the relative velocity.   For shorter wavelengths, the curve moves towards the axes, and the region of validity is smaller.  }
\label{velocityduration}    
\end{figure}

A common approximation used in {\sc sar} imaging schemes assumes that the target is momentarily stationary during the interval in which it interacts with an individual pulse (the so-called ``fast-time'' interval). In this ``start-stop'' approximation, the target is considered to move (relative to the radar platform) only during the ``slow-time'' interval between pulses.   Because the speed of propagation of the pulse is much more rapid than the target motion, the start-stop approximation 
is justified for sufficiently short pulses. In our analysis, however, we deliberately include the case of waveforms whose duration is sufficiently long that the targets and/or platform move appreciably while the data are collected. Figure \ref{velocityduration} shows the approximate regime of validity of the start-stop approximation for for a system operating at $10$ GHz (X-band).  This figure plots $v \leq \lambda / T$, where $v$ is the relative velocity, $\lambda$ is the wavelength at the center frequency ($3$ cm), and $T$ is the waveform duration. For shorter wavelengths, the curve moves towards the axes, and the region of validity is smaller. We see that the start-stop approximation is invalid for high-frequency systems, long-duration waveforms, or high-velocity targets. For example, a target moving 30 m/sec (67 mph) moves 3 mm in 100 $\mu$sec and 3 m in .1 sec. Such distances could easily be comparable to the radar wavelength,  and the resulting errors would cause defocusing effects such as those seen in \cite{Fi,PDF}.
This issue can become especially important when low-power, long-duration waveforms are used.

This paper is an extension of the work \cite{CB}, in which we showed how to combine the temporal, spectral, and spatial attributes of radar data.   In particular, the theory developed below shows how to combine fast-time Doppler and range measurements made from different spatial locations. This approach can be used, for example, for {\sc sar} imaging when relative velocities are large enough so that target echo returns at each look are Doppler-shifted.  Alternatively, this theory shows how to include spatial considerations into classical radar ambiguity theory. In addition, this approach provides a connection between {\sc sar} and Moving Target Indicator ({\sc mti}) radar.  

SAR imaging of moving targets has been previously addressed using the start-stop approximation.  In \cite{Fi}, Fienup analyzed the phase perturbations caused by moving targets and showed how the motion affects the image.   The work \cite{MGZ,HMMS} identifies ambiguities in four-dimensional phase space that result from attempting to image moving targets from a sensor moving along a straight flight path.   The patent \cite{G} and papers \cite{Jao,MD,PDF,P04,ZWXB} all use the start-stop approximation to identify moving targets from radar data.
 The paper \cite{FN} uses a fluid model to impose conservation of mass on a distribution of moving scatterers, and a Kalman tracker to improve the image adaptively.   

The question arises of whether the Doppler shift can be exploited to provide extra information; the theory developed in this paper can be used to address this question.   Some previous work \cite{DTS,K} suggests that Doppler information may be useful for obtaining velocity information about isolated moving targets.  The present paper, on the other hand, handles an arbitrary distribution of moving targets as well as an arbitrary flight path. 

The validity of the start-stop approximation for SAR was previously considered \cite{Tsy} in the case of stationary targets, an antenna moving along a straight line, and chirp waveforms.   This paper can be considered an extension of \cite{Tsy} to the case of moving targets and an arbitrary sequence of waveforms.  

In section 2 we develop a model for monostatic radar data that accommodates temporal, spectral, and spatial diversity.   In section 3  we show how to process the signal to form a phase-space image,  and we express the point-spread function of the imaging system in terms of the ordinary radar ambiguity function.   In section 4 we apply the theory to analyze errors due to  the start-stop approximation.

\section{Model for radar data}
The model for radar data involves a number of ingredients:  the model for wave propagation and scattering, the model for the incident field, the model for the received field, and a number of simplifying assumptions and approximations.  We consider each in turn.  

\subsection{Scalar Wave Propagation}
We model wave propagation with the scalar equation
\begin{equation} \label{scalarwave}
\left( \nabla^2 - {1 \over c^2} {\partial^2 \over \partial t^2} \right) \mathcal E (t, \bm  x) = s(t, \bm x) \, ,
\end{equation}
where $c$ denotes the speed of light in vacuum and $\mathcal E(t,\bm x)$ is one component of the electric field at time $t$ and position $\bm x$.   Here $s$ is a source term that consists of two parts:
 $ s=  s^{\rm in} + s^{\rm sc}$, where $ s^{\rm in}$  models the source due to the transmitting antenna, and  $s^{\rm sc}$ models the effects of target scattering.   The solution $\mathcal E$, which we write as $\mathcal E^{\rm tot}$, therefore splits into two parts: $\mathcal E^{\rm tot} = \mathcal E^{\rm in} + \mathcal E^{\rm sc}$.  The first term, $\mathcal E^{\rm in}$,  satisfies the wave equation for the known, prescribed source $s^{\rm in}$.  This part we call 
the {\em incident} \index{incident field} field --- it is the field in the absence of scatterers. The second part of $\mathcal E^{\rm tot}$ is due to the presence of scattering targets, and this part is called the {\em scattered} field. 

In scattering problems the source term $s^{\rm sc}$ represents the target's {\em response} to an incident field.  This part of the source function will generally depend on the geometric and material properties of the target and on the form and strength of the incident field. Consequently, $s^{\rm sc}$ can be quite complicated to describe analytically. 
Fortunately, for our purposes it is not necessary to provide a detailed analysis of the target's response and we note that for stationary objects consisting of linear materials, we can write $s^{\rm sc}$ as the time-domain convolution
\begin{equation}\label{reflectivity}
s^{\rm sc} (t,\bm{x})=\int q(t-t',\bm{x})\mathcal E^{\rm tot}(t',\bm{x})\d t' , 
\end{equation}
where $q(t,\bm{x})$ is called the reflectivity function. This function is conveniently described in the frequency domain.  We write 
\begin{equation}
Q(\omega, \bm x) = \int \e^{\i\omega t} q(t, \bm x) \d t.
\end{equation}

The frequency-domain reflectivity function $Q(\omega,\bm{x})$ can display a sensitive dependence on $\omega$. When the target is small in comparison with the wavelength of the incident field, for example, $Q\sim\omega^2$ (this behavior is known as ``Rayleigh scattering''). At higher frequencies (shorter wavelengths) the dependence on $\omega$ is typically less pronounced. In the so-called ``optical region'' $Q(\omega,\bm{x})$ is often approximated as being independent of $\omega$ (see, however, \cite{PM}); we use the optical approximation in this paper and simply drop the $\omega$ dependence.  In the time domain, this approximation corresponds to $q(t, \bm z) = \delta (t) Q(\bm z)$.

 If the scatterer moves along the path $\bm x + \bGamma$, where
$\bGamma$  passes through the origin at time $t=0$, then the scatterer reflectivity translates as
\begin{equation}	\label{targetmodel}
q(t, \bm x) = \delta(t) Q(\bm x - \bGamma (t)) .
\end{equation}

In general, both the source and the target will be moving; here we carry out all calculations in the rest frame of the stationary earth.

\subsection{A model for the field from a moving antenna}

We model the antenna as an isotropic point source located at $\bgamma(t)$ (in the earth  reference frame), transmitting (in its own reference frame) the waveform $f(t)$.
The corresponding equation for the time-domain field emanating from the antenna is then 
\begin{equation}	\label{incidenteq}
\nabla^2   \mathcal E^{in} - c_0^{-2} \partial_{tt} \mathcal E^{in} = f(t) \delta(\bm x - \bgamma(t)) ,
\end{equation}
where $c_0$ is the speed of light in vacuum.    
Consequently, the incident field  is \cite{K,CB}
\begin{align}	\label{transmitted}
\mathcal E^{in}(t, \bm x) 
	&=   -\int { \delta(t - t' -|\bm x- \bm y|/c) \over 4 \pi |\bm x - \bm y|} 
		f(t') \delta(\bm y - \bgamma(t')  ) \d t' \d\bm y  \cr
	&= - \int { \delta(t - t' -|\bm x- \bgamma(t')|/c) \over 4 \pi |\bm x - \bgamma(t')|}  f(t') \d t' ,
\end{align}
where we have used the Green's function for a point source \cite{T}
\begin{equation}
g(t, \bm x) = { \delta(t-|\bm x|/c_0) \over 4 \pi |\bm x|}
	= \int { \e^{-\i \omega (t-|\bm x|/c_0)} \over 8 \pi^2 |\bm x|} \d\omega,
\end{equation}
which satisfies
\begin{equation}	\label{pointsource}
\left(\nabla^2 - c_0^{-2} \partial_{tt} \right) g(t,\bm x) = -\delta(t) \delta(\bm x).
\end{equation}
A more realistic antenna beam pattern can easily be included as in \cite{CBbook}.

\subsection{The Scattered Field}
To obtain an equation for the scattered field, we solve \eqref{scalarwave} with the source on the right side given by \eqref{reflectivity}.

\subsubsection{The Lippmann-Schwinger Integral Equation}

We 
use (\ref{transmitted}), with $f$ replaced by $s^{\rm sc}$, to express $\mathcal E^{\rm sc}$ as a
{\em Lippmann-Schwinger} integral equation
\begin{align}	\label{LS}
\mathcal E^{\rm sc}(t,{\bm x}) &=  \iint g(t-\tau, {\bm x}-\bm z) s^{\rm sc}(\tau, \bm z) \d\tau \d\bm z \cr
 &   =\iint g(t-t', {\bm x}-\bm z) Q(\bm z - \bGamma(\tau))
       \mathcal E^{\rm tot} (t', \bm z)  \d t' \d\bm z .
\end{align}

\subsubsection{The Born Approximation}	\label{Bornsection}
For radar imaging, we would like to determine $Q$ from measurements of $\mathcal E^{\rm sc}$ at the antenna.  However, it is difficult to do this directly from \eqref{LS}, because both $Q$ and $\mathcal E^{\rm sc}$ in the neighborhood of the target are unknown, and in \eqref{LS} these unkowns are mutliplied together.  
 This nonlinearity makes it difficult to solve for $Q$.  Consequently, almost all work on radar imaging involves making the {\em Born}\index{Born approximation} approximation,  which is also known as the {\em weak-scattering} or {\em single-scattering}  approximation.  The Born approximation replaces $\mathcal E^{\rm tot}$ on the right side of \eqref{LS} by $\mathcal E^{\rm in}$, which is known.   This results in a formula for $\mathcal E^{\rm sc}$ in terms of $Q$:
 \begin{equation}	\label{Borntime}
\mathcal E^{\rm sc}(t,{\bm x}) \approx \mathcal E^{\rm sc}_B (t, \bm x)
	:=  \iint  g(t-t' , {\bm x}-{\bm z}) Q( {\bm z} - \bGamma(t' )) 
	 \mathcal E^{\rm in}(t'  ,{\bm z})  \d t'  \d{\bm z}.
\end{equation}

In (\ref{Borntime}), we use (\ref{transmitted}) to obtain
\begin{align}\label{Born}
\mathcal E^{\rm sc}_B (t, \bm x) &=
 \int g(t-t', \bm x -  \bm y) Q( \bm y - \bgamma (t'))      
 	\mathcal E^{\rm in} (t',\bm y) \d t' \d^3y \\
	&=  -\int {\delta (t- t' -| \bm x - \bm y|/c) \over 4\pi  |\bm x - \bm y|} 
		\ \int Q( \bm y - \bGamma (t'))  
		 \int { \delta(t' - t'' -|\bm y- \bgamma(t'')|/c) \over 
		 4 \pi |\bm y - \bgamma(t'')|}  f(t'')  \d t'' \d t' \d^3y . 	\nonumber
\end{align}

\subsection{The received field}
Since we are modeling the antenna as an isotropic point source, the 
field received at the antenna, which is located at $\bm x = \bgamma(t)$, is simply the field evaluated at the antenna, $\mathcal E^{\rm sc}_B(t) := \mathcal E^{\rm sc}_B (t, \bgamma(t))$:  
\begin{align}
\mathcal E^{\rm sc}_B (t) = -\int {\delta (t- t' -| \bgamma(t) - \bm y|/c) \over 
	4\pi  |\bgamma(t) - \bm y|} 
		\ \int Q( \bm y - \bGamma (t'))  
		 \int { \delta(t' - t'' -|\bm y- \bgamma(t'')|/c) \over 4 \pi |\bm y - \bgamma(t'')|}  f(t'') \d t'' \d t' \d^3y .
\end{align}

 We now make the change of variables $\bm y \mapsto \bm z = \bm y - \bGamma  (t')$, 
 whose inverse is $\bm y = \bm z + \bGamma(t')$.
This change of variables  converts equation (\ref{Born}) into
\begin{align} 	\label{scatteredfield}
\mathcal E^{\rm sc}_B (t) = -\iiint {\delta (t- t' -| \bgamma(t) - (\bm z + \bGamma(t'))  |/c) \over 
	4\pi  |\bgamma(t) - (\bm z + \bGamma(t')) |} 
		\  Q( \bm z )
		 { \delta(t' - t'' -|\bm z + \bGamma(t') - \bgamma(t'')|/c) \over 4 \pi |\bm z + \bGamma(t') - \bgamma(t'')|}  f(t'') \d t'' \d t' \d^3 z .
\end{align}

Eq. \eqref{scatteredfield} is a model for the field from a single localized moving target.   The right side of \eqref{scatteredfield} can be interpreted as follows.  That part of the waveform $f$ that is transmitted at time $t''$ from location $\bgamma(t'')$  travels to the target, arriving at time $t'$.  At time $t'$, the target that started at $\bm z$ is now at location $\bm z + \bGamma(t')$.  The wave scatters with relative strength $Q(\bm z)$, and then propagates to the receiver, arriving at time $t$.  At time $t$, the receiver is at position $\bgamma(t)$.  

In the special case of a stationary reflector of unit strength positioned at the origin $\bm z = \bm 0$,
we see that the times $t''$, $t'$, and $t$ are related by
\begin{align}	\label{centertimes}
0 &=  t' - t'' - |\bgamma(t'')|/c\cr
0 &= t - t' - |\bgamma(t)|/c .
\end{align}

If the scene involves a distribution of targets with reflectivities $Q_{\bGamma}$, each moving on a different path $\bGamma$, then  $\mathcal E^{\rm sc}_B$
is integrated over all these  target paths.  
To avoid dealing with such a path integral, we consider only the following simpler case.

\subsection{The Case of Pulsed Radar}	\label{pulses}
For the rest of the paper, we  assume that the waveform is a train of pulses and that the platform and target motion is slow relative to the speed of light.    

In particular, we take $f$ to be a train of pulses of the form
\begin{equation}\label{train}
f(t)=\sum_m f_m(t-T_m^T) , 
\end{equation}
where  the delay between successive pulses is sufficiently large so that  successive pulses do not overlap.  

 A pulsed system allows us to introduce a ``slow time" variable, namely the pulse number $m$.   We reference the times $t$, $t'$, and $t''$ to the following (``fast") times associated with the $m$th pulse.   
We write  $t' = T_m'$ and $t= T_m^R$ for the solutions to (\ref{centertimes}) corresponding to $t'' = T_m^T$, so that
\begin{align}	\label{centerTimes}
0 &=  T_m' - T_m^T - |\bgamma(T_m^T)|/c\cr
0 &= T_m^R - T_m' - |\bgamma(T_m^R)|/c .
\end{align}
If the time intervals $T_{m+1}^T - T_m^T$ are sufficiently short and the motion is smooth, we can make the expansions
\begin{align}
\bgamma(t'') &= \bgamma(T_m^T) + \dot{\bgamma}(T_m^T) (t'' - T_m^T) + \cdots \cr
\bgamma(t) &= \bgamma(T_m^R) + \dot{\bgamma}(T_m^R) (t - T_m^R) + \cdots \cr
\bGamma(t') &= \bGamma(T_m') + \dot{\bGamma}(T_m') (t' - T_m') + \cdots .
\end{align}

 In making a distinction between the expansion times  $T^T_m, T^R_m$ and $T'_m$, we are including the case in which the platform is extremely distant from the target.  A simpler version of the theory would be obtained by replacing the expansion times $T^R_m $ and $T'_m$ by $T^T_m$.  

\subsection{Spotlight SAR}	\label{spotlight}
We now assume that  the distance from the origin to the sensor position $\bgamma$ is much larger than the distance from the origin to the target, and also much larger than the distance travelled by the target or sensor during any of the time intervals $T_{m+1}^T - T_m^T$.   With these assumptions we can make the following expansions:
\begin{align}	\label{smallslowscene}
|\bm z + \bGamma(t') - \bgamma(t)  | &=  |\bm z + \bGamma(T_m') +
		 \dot{\bGamma}(T_m') (t' - T_m')  -  \bgamma(T_m^R) -
		 \dot{\bgamma}(T_m^R) (t - T_m^R)  + \cdots |   \cr
&= |\bgamma(T_m^R)| - \widehat{\bgamma}(T_m^R) \bdot \left[ \bm z + \bGamma(T_m') +
		 \dot{\bGamma}  (T_m') (t' - T_m') -  \dot{\bgamma}(T_m^R) (t - T_m^R)  \right] + \cdots  \cr
|\bm z + \bGamma(t') - \bgamma(t'')| &=  |\bm z + \bGamma(T_m') +
		 \dot{\bGamma}(T_m') (t' - T_m')  - \bgamma(T_m^T) 
		 - \dot{\bgamma}(T_m^T) (t'' - T_m^T) + \cdots | \cr
&= |\bgamma(T_m^T)| - \widehat{\bgamma}(T_m^T) \bdot \left[ \bm z + \bGamma(T_m') +
		 \dot{\bGamma}  (T_m') (t' - T_m') -  \dot{\bgamma}(T_m^T) (t'' - T_m^T)  \right] + \cdots  .
\end{align}
Below we use the notation $\bgamma^T_m = \bgamma(T_m^T)$ ,  $\bgamma^R_m = \bgamma(T_m^R)$,
 $\bGamma_m = \bGamma(T_m')$ , and $\bm v_m = \dot{\bGamma}(T_m')$.   
 
 To carry out the $t''$ integration of (\ref{scatteredfield}), we use the approximations 
(\ref{smallslowscene}) together with  the change of variables
\begin{equation}
t'' \mapsto \tilde{t}'' =  t' - t'' -|\bm z + \bGamma(t') - \bgamma(t'')|/c
\end{equation}
whose Jacobian is 
\begin{equation}
\left| {\partial t'' \over \partial \tilde{t}'' } \right| 	
	= {1 \over  \left| \partial \tilde{t}''/\partial t'' \right|  }  
	\approx \left| {1   \over  
	- 1 - \widehat{\bgamma}^T_m \bdot \dot{\bgamma}^T_m /c } \right| .
\end{equation}
The $t''$ integration of (\ref{scatteredfield}) contributes only when $\tilde{t}'' = 0$, which occurs at a value of $t''$ that we shall denote by $\tau_m''(t', \bm z, \bm v_m)$,  which is given approximately by
\begin{align}
\tau_m''(t', \bm z, \bm v_m) \approx {1 \over  1 + \widehat{\bgamma}^T_m \bdot \dot{\bgamma}^T_m/c}
	\left[  \left( 1 + \widehat{\bgamma}^T_m \bdot \bm v_m /c \right) t' 
	- |\bgamma^T_m|/c + \widehat{\bgamma}^T_m \bdot \left( 
	\bm z + \bGamma_m - \bm v_m T_m' + \dot{\bgamma}^T_m T_m^T \right) /c
	\right] .
\end{align}
The time $\tau_m''(t', \bm z)$ is the {\em retarded time}, that is, the time when the field interacting with the target at time $t'$ was transmitted.  

In (\ref{scatteredfield}) we also carry out the $t'$ integration by making the change of variables
\begin{equation}
t' \mapsto \tilde{t}' = t- t' -|\bm z + \bGamma(t') - \bgamma(t)    |/c
\end{equation}
whose Jacobian is
\begin{equation}
\left| {\partial t' \over \partial \tilde{t}'} \right| = {1 \over \left| \partial \tilde{t}'/\partial t' \right| } 
	\approx  \left| {1 \over -1 +  \widehat{\bgamma}^R_m \bdot  \bm v_m /c } \right| .
\end{equation}
The delta function is supported only at $\tilde{t}'=0$, which occurs at a value
$t' = \tau'_m(t, \bm z, \bm v_m)$ defined by
\begin{equation}	\label{t'def}
0 = t- t' - |\bm z + \bGamma(t') - \bgamma(t)    |/c .
\end{equation}
With the approximations (\ref{smallslowscene}), we obtain
\begin{equation}
\tau'_m(t, \bm z, \bm v_m) \approx  {1 \over 1 -  \widehat{\bgamma}^R_m \bdot \bm v_m /c}
	\left[  \left( 1 - \widehat{\bgamma}^R_m \bdot \dot{\bgamma}^R_m /c \right) t
	- |\bgamma^R_m|/c + \widehat{\bgamma}^R_m \bdot \left( 
	\bm z + \bGamma_m - \bm v_m T_m' + \dot{\bgamma}^R_m T_m^R \right) /c
	\right] .
\end{equation}

The expression (\ref{scatteredfield}) then becomes
\begin{align}	\label{scatteredfield2}
\mathcal E^{\rm sc}_B (t) &= -\sum_m \int 
	 { f_m (\phi_m(t, \bm z, \bm v_m) )  \  Q( \bm z ) \over 
	(4\pi)^2  |\bgamma^T_m| |\bgamma^R_m | (1 - \widehat{\bgamma}^R_m \bdot \bm v_m /c)
	(1 + \beta^T_m) }  \d^3 z ,
\end{align}
where 
\begin{align}	\label{dataphase}
\phi_m(t, \bm z, \bm v_m) &:= \tau_m'' (\tau'_m(t, \bm z, \bm v_m) , \bm z, \bm v_m)  -T_m^T \cr
& \approx 
	{1 \over 1 + \beta^T_m} \left[ \alpha_{\bm v,m} \left( [1 - \beta^R_m] t - 
	R_m^R(\bm z, \bm v_m) /c \right) - R_m^T(\bm z, \bm v_m) /c \right]  -T_m^T 
\end{align}
and
\begin{align}	\label{notation}
 \beta^T_m &= \widehat{\bgamma}^T_m \bdot \dot{\bgamma}^T_m/c \cr
 \beta^R_m &=  \widehat{\bgamma}^R_m \bdot \dot{\bgamma}^R_m /c \cr
 \alpha_{\bm v_m, m} & = {1 + \widehat{\bgamma}^T_m \bdot \bm v_m /c  \over
 	1 -  \widehat{\bgamma}^R_m \bdot \bm v_m /c}  \cr
R^R_m(\bm z, \bm v_m) & = |\bgamma^R_m| - \widehat{\bgamma}^R_m \bdot \left( 
	\bm z + \bGamma_m - \bm v_m T_m' + \dot{\bgamma}^R_m T_m^R \right)    \cr
 R^T_m(\bm z, \bm v_m)  &= |\bgamma^T_m| - \widehat{\bgamma}^T_m \bdot \left( 
	\bm z + \bGamma_m - \bm v_m T_m' + \dot{\bgamma}^T_m T_m^T \right) .
\end{align}
The quantities $\beta^T$ and $\beta^R$ are determined by the {\em squint} angle (angle relative to broadside) of the antenna while it is transmitting and receiving, respectively, and $\alpha$ is the {\em Doppler scale factor}.

We assume that
the antenna illuminates a limited region of the earth and that
all targets are contained in a bounded region in phase space.   Moreover, we assume that the pulse repetition intervals are chosen so that the response from pulse $m$ can be isolated from that of other pulses.   
 Thus the data due to the $m$th pulse is just one term of \eqref{scatteredfield2}:
\begin{align}	\label{scatteredfield3}
\mathcal E^{\rm sc}_m (t) &= - \int 
	 { f_m (\phi_m( t, \bm z, \bm v_m))  \  Q( \bm z ) \over 
	(4\pi)^2  |\bgamma^T_m| |\bgamma^R_m| (1 - \widehat{\bgamma}^R_m \bdot \bm v_m /c)
	(1 + \beta^T_m) }  \d^3 z .
\end{align}

Expression (\ref{scatteredfield3}) applies to moving targets whose velocity during pulse $m$ is  $\bm v_m$.  To include multiple targets that may be moving with different velocities, we integrate over $\bm v_m$.  
\begin{align}	\label{data}
\mathcal E^{\rm sc}_m (t) &= - \int 
	 { f_m (\phi_m( t, \bm z, \bm v_m) )   \  Q( \bm z, \bm v_m ) \over 
	(4\pi)^2  |\bgamma^T_m| |\bgamma^R_m| (1 - \widehat{\bgamma}^R_m \bdot \bm v_m /c)
	(1 + \beta^T_m) }  \d^3 z \d^3 v_m .
\end{align}

\subsection{Example:  Data model for circular {\sc sar} and constant-velocity targets}
For the case of Circular {\sc sar} with no squint,  we have 
$\widehat{\bgamma}^R_m \bdot \dot{\bgamma}^R_m = 0 =
\widehat{\bgamma}^T_m \bdot  \dot{\bgamma}^T_m$, which implies that (\ref{notation}) becomes 
\begin{align}
\beta_m^T &= 0 = \beta_m^R \cr
R^{R,\circ}_m(\bm z, \bm v_m) & =R- \widehat{\bgamma}^R_m \bdot \left( 
	\bm z + \bGamma_m - \bm v_m T_m'  \right)   
	=  R- \widehat{\bgamma}^R_m \bdot \bm z\cr
 R^{T,\circ}_m(\bm z, \bm v_m)  &= R - \widehat{\bgamma}^T_m \bdot \left( 
	\bm z + \bGamma_m - \bm v_m T_m' \right) 
	= R - \widehat{\bgamma}^T_m \bdot \bm z ,
\end{align}
where we have used the assumption that target velocity is constant to write $\bGamma_m = \bm v_m T_m'  $. 
Thus (\ref{data}) with (\ref{dataphase}) becomes
\begin{align}	\label{circulardata}
\mathcal E^{\circ}_m (t) &= -{1 \over (4\pi)^2 R^2  }  \int 
	  f_m (\phi_m^\circ(t, \bm z, \bm v_m) )   \  Q( \bm z, \bm v_m ) 
	 \d^3 z \d^3 v_m ,
\end{align}
where from (\ref{dataphase}), the phase is
\begin{align}	\label{circularphase}
\phi_m^\circ(t, \bm z, \bm v_m) &=
	 \alpha_{\bm v,m} \left(   t - 
	R^{R, \circ}_m(\bm z, \bm v_m) /c \right) - R^{T,\circ}_m(\bm z, \bm v_m) /c - T_m^T  \cr
&\approx \left(1 + {\bm B_m \bdot \bm v_m \over c} \right)  
\left[ t  - {R\over c}
	   + \widehat{\bgamma}^R_m \bdot
	  \left( { \bm z  \over c} \right)  \right] 
	- \left[ {R\over c} - \widehat{\bgamma}^T_m \bdot
	\left( { \bm z  \over c} \right)  \right]   -T_m^T \cr
&=  t - T_m^T - 2{R\over c} + {\bm B_m \over c}  \bdot \left[\bm z  
	+ \bm v_m  \left( t - {R \over c}  
		-  \widehat{\bgamma}^R_m \bdot { \bm z  \over c} \right)  \right] .
\end{align}
where $\bm B_m =  \widehat{\bgamma}^T_m + \widehat{\bgamma}^R_m$ is the ``bistatic bisector".   
For a system in which $\bgamma^T_m \approx \bgamma^R_m$, $\bm B_m$ is simply the look direction.  

%

\section{Image Formation}

We process the data for each pulse, and then we combine the results over the pulse number $m$.

\subsection{Signal Processing at each Pulse}

To process the data \eqref{data} at each pulse $m$, we apply a weighted matched filter, {\it i.e.,} we correlate the scattered field \eqref{data} with a weighted version of the signal  we {\em expect} to see from each point $\bm p$ and each velocity $\bm u$.  This results in   
\begin{align}	\label{mimage}
I_m( \bm p, \bm u) & = -\int f_m^* \left( \phi_m(t, \bm p, \bm u_m)  \right) 
	 (4 \pi)^2 |\bgamma^T_m| |\bgamma^R_m| (1 - \widehat{\bgamma}^R_m \bdot \bm v_m /c)
	(1 + \beta^T_m)
	\mathcal E^{\rm sc}_m(t) \d t .
\end{align}

\subsection{The Impulse-Response Function}
The impulse-response function for each pulse is obtained by substituting (\ref{data}) into (\ref{mimage}).  This results in
\begin{equation}	\label{imagem}
I_m(\bm p, \bm u_m) = \iint K_m(\bm u_m, \bm p; \bm z, \bm v_m ) Q( \bm z, \bm v_m )  
	\d^3 v_m    \d^3 z ,
\end{equation}
where
\begin{align}	\label{psf1}
K_m(\bm u_m , \bm p; \bm z, \bm v_m )  
	=\int f_m^* \left( \phi_m(t, \bm p, \bm u_m)   \right) 
	   f_m(\phi_m(t, \bm z, \bm v_m)) 
	{   1 - \widehat{\bgamma}^R_m \bdot \bm u_m /c  \over 
	 1 - \widehat{\bgamma}^R_m \bdot \bm v_m /c }   \d t	 .
\end{align}

In (\ref{psf1}) we make the change of variables
\begin{equation}
t \mapsto t' = \phi_m(t, \bm z, \bm v_m)
\end{equation}
for which the Jacobian is
\begin{equation}
\left| {\partial t \over \partial t'} \right| = {1 \over \left| {\partial t' \over \partial t} \right| } 
	= \left| {1 + \beta^T_m \over
	\alpha_{\bm v, m} (1 - \beta^R_m)}  \right| .
\end{equation}
This converts (\ref{psf1}) into
\begin{align}	\label{psfm}
K_m(\bm u_m , \bm p; \bm z, \bm v_m )  =   \mathcal A_m  \left( 
	{\alpha_{\bm u_m , m} \over \alpha_{\bm v_m, m} }  ,
	\Delta \tau(\bm p,  \bm u;  \bm z, \bm v) \right)  
	{   1 - \widehat{\bgamma}^R_m \bdot \bm u_m /c  \over 
	 1 - \widehat{\bgamma}^R_m \bdot \bm v_m /c } 
	\   \left| {1 + \beta^T_m \over
	\alpha_{\bm v, m} (1 - \beta^R_m)}  \right|  ,
\end{align}
where
\begin{equation}\label{WBambiguity}
\mathcal A_m  (\sigma, \tau) = \int f_m^* (\sigma[t - \tau]) f_m(t) \d t 
\end{equation}
is the wideband {\em radar ambiguity function} \cite{Sw} and where
\begin{align}	\label{Deltatau}
\Delta \tau_m(\bm p,  \bm u_m;  \bm z, \bm v_m) &=  
	{\alpha_{\bm v_m, m} \over  (1 + \beta^T_m)} \bigg[ 
	{R^T_m(\bm p, \bm u) \over c \alpha_{\bm u, m}}
	- {R^T_m(\bm z, \bm v_m) \over  c \alpha_{\bm v, m}}
	+ {R^R_m(\bm p, \bm u) \over c}  - { R^R_m(\bm z, \bm v) \over c} \cr
& \qquad 	+ (1 + \beta^T_m) T_m^T  \left( {1 \over  \alpha_{\bm u, m} }
		-{ 1 \over  \alpha_{\bm v, m} }  \right)	\bigg] .
\end{align}

For the case of a stationary radar,  \eqref{imagem} and \eqref{psfm} express the well-known fact \cite{L} that a fixed radar can form a range-Doppler image of moving targets, and the system's ability to obtain range and velocity information is determined by the transmitted waveform $f_m$ via the radar ambiguity function.  What is new here is that we have carefully accounted for the motion and spatial positions of the antenna and target distribution.  

\subsection{Phase-Space Image Formation and Point-Spread Function}

We now have an image at each pulse, and a corresponding impulse-response function that depends on the pulse number $m$. 
This information could, for example, be fed into a tracking algorithm.   In the present paper, we
consider instead the formation of a phase-space image; in other words, we want to obtain information about both position and velocity of the targets.  In order to combine information from multiple pulses, we make the assumption that the target velocity does not vary with $m$.
We form the phase-space image as a weighted sum \eqref{imagem} over $m$, obtaining
\begin{align}	\label{image}
I (\bm p, \bm u) = \sum_m I_m(\bm p, \bm u) \left| {\alpha_{\bm u, m} (1 - \beta^R_m)
	 \over  1 + \beta^T_m }  \right| 
	=\sum_m \int   K_m(\bm p,  \bm u;  \bm z, \bm v)  
		\left| {\alpha_{\bm p, m} (1 - \beta^R_m)	 \over  1 + \beta^T_m }  \right| 
	Q(\bm z, \bm v)   \d^3 z  \d^3 v  .
\end{align}
We write this as
\begin{equation}
I (\bm p, \bm u)
=  \int K(\bm p,  \bm u;  \bm z, \bm v) Q(\bm z, \bm v)   \d^3 z  \d^3 v  ,
\end{equation}
where the full point-spread function $K$ is the sum over $m$ 
of the impulse-response functions \eqref{psf1}:
\begin{equation}	\label{fullpsf}
K(\bm p,  \bm u;  \bm z, \bm v)  = \sum_m  K_m(\bm p,  \bm u;  \bm z, \bm v) 
	\left| {\alpha_{\bm u, m} (1 - \beta^R_m)
	 \over  1 + \beta^T_m }  \right|  .
\end{equation}

We write the point-spread function   (\ref{fullpsf}) as
\begin{equation}
\boxed{
\label{fullpsf2}
K(\bm p,  \bm u;  \bm z, \bm v)  
	= \sum_m  \left| {\alpha_{\bm u, m} \over \alpha_{\bm v, m} } \right|   \mathcal A_m 
	 \left( 
	{\alpha_{\bm u , m} \over \alpha_{\bm v, m} }  ,
	\Delta \tau_m (\bm p,  \bm u;  \bm z, \bm v) \right)  
	{   1 - \widehat{\bgamma}^R_m \bdot \bm u /c  \over 
	 1 - \widehat{\bgamma}^R_m \bdot \bm v /c } .  
}
\end{equation}
 
 Thus we see that {\it the phase-space point-spread function (PSF) is a weighted, coherent sum of ordinary radar ambiguity functions, evaluated at arguments that depend on the difference in positions and velocities.}
  
 This theory can potentially be applied to a variety of {\sc sar} image analysis problems involving moving targets, including the design of waveforms for tracking, analysis of resolution in phase space, etc.     Below we show one example, namely an analysis of the validity of the start-stop approximation. 


\section{Application:  Validity of the Start-Stop Model}
Almost all {\sc sar} signal processing and analysis makes use of the {\em start-stop} approximation, also called the {\em stop-and-shoot} approximation, which  assumes that neither target nor antenna is moving while each interacts with the wave.

\subsection{The Start-Stop Model for Spotlight SAR Data}

The usual start-stop model for monostatic SAR is
\begin{equation}
\mathcal E^{\rm sc}_m (t) = \int \int {f_m \left( t - 2R_m(\bm z, \bm v) /c   \right) \over  
	(4\pi)^2  |\bgamma_m | |\bgamma_m |}  Q(\bm z, \bm v_m) \d^3 z \d^3 v_m ,
\end{equation}
where $R_m$ denotes the distance from the stationary antenna to the target at $\bm z$ moving with velocity $\bm v$. 

Below we use the theory developed above to extend the start-stop model to include the case of a fast-moving antenna  very distant from the scene, in which case the antenna  may not be at the same location when it transmits and receives.  
In this case, the start-stop approximation is to assume that the antenna is stationary while it is transmitting and while it is receiving, but it may move between transmission and reception.  

Under the start-stop assumption, (\ref{smallslowscene}) is instead
\begin{align}	\label{startstopTaylor}
|\bm z + \bGamma(t') - \bgamma(t)  | &\approx  |\bm z + \bGamma(T_m') 
		  -  \bgamma(T_m^R)  |   
\approx |\bgamma(T_m^R)| - \widehat{\bgamma}(T_m^R) \bdot \left[ \bm z + \bGamma(T_m')  \right]
	+ \cdots \cr
|\bm z + \bGamma(t') - \bgamma(t'')| &\approx  |\bm z + \bGamma(T_m') 
		  - \bgamma(T_m^T)  | 
\approx |\bgamma(T_m^T)| - \widehat{\bgamma}(T_m^T) \bdot \left[ \bm z + \bGamma(T_m') \right] 
	+ \cdots .
\end{align}
Thus, using $\bGamma_m = \bm v_m T'_m$,  the start-stop signal model is 
\begin{align} 	\label{startstopfield}
\mathcal E^{\rm sc}_{ss} (t) &= -\iiint {\delta (t- t' -| \bgamma^R_m|/c  + \widehat{\bgamma}^R_m \bdot \left[ \bm z + \bm v_m T'_m \right] /c) \over 
	4\pi  |\bgamma^R_m |} 	\  Q( \bm z ) \cr
& \qquad 		 { \delta(t' - t'' -|\bgamma^T_m|/c + \widehat{\bgamma}^T_m \bdot \left[ \bm z + \bm v_m T'_m \right] /c) \over 4 \pi |\bgamma^T_m|}  f_m(t'' -T_m^T ) \d t'' \d t' \d^3 z \cr
&= \int {f_m \left( t - (R^{T,ss}_m(\bm z, \bm v) + R^{R,ss}_m(\bm z, \bm v)  )/c -T_m^T  \right) \over  
	(4\pi)^2  |\bgamma^R_m | |\bgamma^T_m |}  Q(\bm z, \bm v_m) \d^3 z \d^3 v_m ,
\end{align}
where
\begin{align}
R^{T,ss}_m(\bm z, \bm v) &=  |\bgamma^T_m| - \widehat{\bgamma}^T_m \bdot \left[ \bm z + \bm v T'_m \right] \cr
R^{R,ss}_m(\bm z, \bm v) &= | \bgamma^R_m|  - \widehat{\bgamma}^R_m \bdot \left[ \bm z + \bm v T'_m \right] .
\end{align}
 Here we have used the same definitions for $\bgamma^T_m$, $\bgamma^R_m$, $T'_m$, etc., as in sections \ref{pulses} and \ref{spotlight}. 

\subsection{Start-Stop Image Formation for Spotlight SAR}
The image is formed as
\begin{align}
I_{ss}(\bm p, \bm u) = \sum_m (4\pi)^2  |\bgamma^R_m | |\bgamma^T_m |
	\int f^*_m \left( t -T_m^T  - (R^{T,ss}_m(\bm p, \bm u) + R^{R,ss}_m((\bm p, \bm u) )/c \right) 
	 \mathcal E^{\rm sc}_{ss} (t) \d t ,
\end{align}
which gives rise to a PSF of the form
\begin{align}
K_{ss}(\bm p, \bm u, \bm z, \bm v) 
&= \sum_m  \int f^*_m \left( t -T_m^T  - (R^{T,ss}_m(\bm p, \bm u) + R^{R,ss}_m((\bm p, \bm u) )/c \right)  \cr
& \qquad 	f_m \left(t -T_m^T   - (R^{T,ss}_m(\bm z, \bm v) + R^{R,ss}_m(\bm z, \bm v)  )/c   \right)  \d t \cr
&= \sum_m  \mathcal A_m (1, \Delta \tau_m^{ss})  ,
\end{align}
where 
\begin{align}
\Delta \tau_m^{ss} =   (R^{T,ss}_m(\bm z, \bm v) + R^{R,ss}_m(\bm z, \bm v)  )/c  
	- (R^{T,ss}_m(\bm p, \bm u) + R^{R,ss}_m((\bm p, \bm u) )/c 
\end{align}
is the difference in travel times between the antenna and targets whose phase-space coordinates are
$(\bm z, \bm v)$ and $(\bm p, \bm u)$. 

A focussed image will be obtained when
\begin{align}	\label{startstopfocus}
0 &= (R^{T,ss}_m(\bm z, \bm v) + R^{R,ss}_m(\bm z, \bm v)  )
	- (R^{T,ss}_m(\bm p, \bm u) + R^{R,ss}_m((\bm p, \bm u) )\cr
&= 	|\bgamma^T_m|/c - \widehat{\bgamma}^T_m \bdot \left[ \bm z + \bm v T'_m \right] 
	- \left( |\bgamma^T_m|/c - \widehat{\bgamma}^T_m \bdot \left[ \bm p + \bm u T'_m \right] \right) \cr
& \qquad 
	+ | \bgamma^R_m|/c  - \widehat{\bgamma}^R_m \bdot \left[ \bm z + \bm v T'_m \right]
	- \left( | \bgamma^R_m|/c  - \widehat{\bgamma}^R_m \bdot \left[ \bm p + \bm u T'_m \right] \right) \cr 
&= \left(  \widehat{\bgamma}^T_m + \widehat{\bgamma}^R_m\right) 
		 \bdot \left[ ( \bm p - \bm z) + (\bm u - \bm v) T'_m \right] 
= \bm B_m \bdot \left[ ( \bm p - \bm z) + (\bm u - \bm v) T'_m \right] ,
\end{align}
where we have written
\begin{equation}
\bm B_m = \widehat{\bgamma}^T_m  + \widehat{\bgamma}^R_m .
\end{equation}

\subsection{Mismatched Processing}
If we incorrectly believe the start-stop data model to be accurate, we will form an image via the same processing as for the start-stop model, namely
\begin{align}	\label{mismatchimage}
I_{\times}(\bm p, \bm u) = \sum_m (4\pi)^2  |\bgamma^R_m | |\bgamma^T_m |
	\int f^*_m \left( t -T_m^T  - {R^{T,ss}_m(\bm p, \bm u) + R^{R,ss}_m((\bm p, \bm u) \over c}  \right) 
	 \mathcal E^{\rm sc}_{m} (t) \d t ,
\end{align}
where $ \mathcal E^{\rm sc}_{m} (t)$ is given by \eqref{data}.   Substituting \eqref{data}  for $\mathcal E^{\rm sc}_{m} (t)$ in 
\eqref{mismatchimage}, we obtain
\begin{align}
I_{\times}(\bm p, \bm u) &= - \sum_m 
	\int f^*_m \left( t -T_m^T  - {R^{T,ss}_m(\bm p, \bm u) + R^{R,ss}_m((\bm p, \bm u) \over c}  \right)  \cr
&\qquad 	   \int 
	 { f_m (\phi_m( t, \bm z, \bm v_m) )   \  Q( \bm z, \bm v_m ) \over 
	 (1 - \widehat{\bgamma}^R_m \bdot \bm v_m /c)
	(1 + \beta^T_m) }  \d^3 z \d^3 v_m\d t .
\end{align}
The mismatched PSF is
\begin{align}	\label{mismatchPSF}
K_{\times}(\bm p, \bm u; \bm z, \bm v) &=  \sum_m 
	\int f^*_m \left( t -T_m^T  - {R^{T,ss}_m(\bm p, \bm u) + R^{R,ss}_m((\bm p, \bm u) \over c}  \right)  
	  { f_m (\phi_m( t, \bm z, \bm v_m) )  \over 
	 (1 - \widehat{\bgamma}^R_m \bdot \bm v_m /c)
	(1 + \beta^T_m) }  \d t .
\end{align}
In \eqref{mismatchPSF}, we make the change of variables $t \mapsto t' = \phi_m( t, \bm z, \bm v_m)$, whose inverse is
\begin{equation}
t = {1 \over 1 - \beta^R_m } \left( {1 + \beta^T_m \over \alpha_{\bm v_m, m}} (t' + T^T_m)
	+ {R^T_m(\bm z, \bm v_m) \over c  \ \alpha_{\bm v_m, m}}  
	+ {R^R_m(\bm z, \bm v) \over c} \right) 
\end{equation} 
and whose Jacobian is 
\begin{equation}
{\partial t \over \partial t'} = {1 + \beta^T_m \over ( 1 - \beta^R_m )\alpha_{\bm v_m, m}} .
\end{equation}
This change of variables converts \eqref{mismatchPSF} to
\begin{align}
K_{\times}(\bm p, \bm u; \bm z, \bm v) &=  \sum_m 
	\int f^*_m \left( {1 + \beta^T_m \over ( 1 - \beta^R_m )\alpha_{\bm v_m, m}} 
	(t' - \Delta \tau^\times_m) \right)    { f_m (t' )  \over 
	 (1 + \widehat{\bgamma}^T_m \bdot \bm v_m /c)
	 ( 1 - \beta^R_m )\ }  \d t' \cr
&= \sum_m \mathcal A \left( {1 + \beta^T_m \over ( 1 - \beta^R_m )\alpha_{\bm v_m, m}} , 
		\Delta \tau^\times_m \right) { 1  \over 
	 (1 + \widehat{\bgamma}^T_m \bdot \bm v_m /c)
	 ( 1 - \beta^R_m ) } ,
\end{align}
where
\begin{align}
\Delta \tau^\times_m & = 
	\bigg( T^T_m  \left[ {(1 - \beta^R_m)  \alpha_{\bm v_m, m} \over 1 + \beta^T_m}   -1 \right]
	+  {R^{T,ss}_m(\bm p, \bm u)   (1 - \beta^R_m)  \alpha_{\bm v_m, m} \over (1 + \beta^T_m) c} 	
	- { R^T_m(\bm z, \bm v_m) \over   (1 + \beta^T_m) c  }  \cr
&	+ { R^{R,ss}_m(\bm p, \bm u) (1 - \beta^R_m)  \alpha_{\bm v_m, m} \over (1 + \beta^T_m) c }
	 -  {  R^R_m(\bm z, \bm v) \alpha_{\bm v_m, m} \over (1 + \beta^T_m)  c} 
	\bigg).
\end{align}

A focused image is the set of $(\bm p, \bm u; \bm z, \bm v)$ that are at the ambiguity function peak for all $m$:
\begin{align}	\label{ambiguitypeak}
1 &= {1 + \beta^T_m \over ( 1 - \beta^R_m )\alpha_{\bm v_m, m}} \cr
0 & = \Delta \tau^\times_m .
\end{align}
 The first line of \eqref{ambiguitypeak} is the ``Doppler" condition, and the second line is the ``range" condition.
We note that the reciprocal of the expression on the first line of  \eqref{ambiguitypeak} can be written
\begin{align}
{(1 - \beta^R_m)  \alpha_{\bm v , m} \over 1 + \beta^T_m} 
&= {1 - \widehat{\bgamma}^R_m \bdot \dot{\bgamma}^R_m /c \over
		1+ \widehat{\bgamma}^T_m \bdot \dot{\bgamma}^T_m/c }\ 
		{1 + \widehat{\bgamma}^T_m \bdot \bm v /c  \over
 	1 -  \widehat{\bgamma}^R_m \bdot \bm v /c}   
\approx \left( 1 - \widehat{\bgamma}^R_m \bdot \dot{\bgamma}^R_m /c
	 -  \widehat{\bgamma}^T_m \bdot \dot{\bgamma}^T_m/c \right)
	 \left( 1 +\bm B_m \bdot \bm v/c  \right) \cr
&= 1 - \widehat{\bgamma}^R_m \bdot \dot{\bgamma}^R_m /c
	 -  \widehat{\bgamma}^T_m \bdot \dot{\bgamma}^T_m/c +\bm B_m \bdot \bm v/c  + \cdots,
\end{align}
which, for the case of constant platform velocity
($\dot{\bgamma}:= \dot{\bgamma}^R_m = \dot{\bgamma}^T_m$), is
\begin{equation}
{(1 - \beta^R_m)  \alpha_{\bm v , m} \over 1 + \beta^T_m}  
\approx  1 + \bm B_m \bdot (\bm v - \dot{\bgamma}) /c  + \cdots.
\end{equation}

\subsection{Error Analysis for Ideal High-Range-Resolution Waveforms}
We expect that the start-stop approximation will be best in the case of a waveform whose ambiguity function is insensitive to its first argument, {\it i.e.,} for a high-range-resolution waveform such as a very short pulse.
For such an ideal high-range-resolution waveform, a focused image is attained when
\begin{align}	\label{HRRfocus}
0= \Delta \tau^\times_m 
& =  T^T_m  \left[ {(1 - \beta^R_m)  \alpha_{\bm v , m} \over 1 + \beta^T_m}   -1 \right]
	+  \left( R^{T,ss}_m(\bm p, \bm u) + R^{R,ss}_m(\bm p, \bm u) \right)   
	 {(1 - \beta^R_m)  \alpha_{\bm v, m} \over (1 + \beta^T_m) c} 	\cr
&
	- \left(R^T_m(\bm z, \bm v_m) +   R^R_m(\bm z, \bm v) \alpha_{\bm v_m, m} \right) 
	{ 1 \over   (1 + \beta^T_m) c  }  .
\end{align}
For the case of constant platform velocity,  \eqref{HRRfocus} becomes
\begin{align}	\label{HRRfocus1}
0&= T^T_m  \left[ {(1 - \beta^R_m)  \alpha_{\bm v, m} \over 1 + \beta^T_m}   -1 \right]
	+ \left( |\bgamma^T_m| - \widehat{\bgamma}^T_m \bdot \left[ \bm p + \bm u T'_m \right] 
	+ | \bgamma^R_m|  - \widehat{\bgamma}^R_m \bdot \left[ \bm p + \bm u T'_m \right] \right) 
	{(1 - \beta^R_m)  \alpha_{\bm v, m} \over (1 + \beta^T_m) c} \cr
& \qquad - {1 \over   (1 + \beta^T_m) c  }  
	\bigg[ |\bgamma^T_m| - \widehat{\bgamma}^T_m \bdot \left( 
	\bm z + \dot{\bgamma}^T_m T_m^T \right) 
	+   \left[ |\bgamma^R_m| - \widehat{\bgamma}^R_m \bdot \left( 
	\bm z +  \dot{\bgamma}^R_m T_m^R \right)  \right]  
	\alpha_{\bm v, m} \bigg] \cr
&\approx  (T'_m - |\bgamma^T_m|/c) \bm B_m \bdot (\bm v - \dot{\bgamma} )/c
	+ \left( R^{RT}_m   - \bm B_m \bdot \left[ \bm p + \bm u T'_m \right] \right)
	\left( 1 + \bm B_m \bdot (\bm v - \dot{\bgamma} )/c \right)/c  \cr
& \qquad - \left( 1 - \widehat{\bgamma}^T_m \bdot \dot{\bgamma}^T_m/c \right) 
	\bigg[ R^{RT}_m - \bm B_m \bdot  \bm z 
	- \widehat{\bgamma}^T_m \bdot \dot{\bgamma}^T_m (T'_m - |\bgamma^T_m|/c)
	- \widehat{\bgamma}^R_m \bdot  \dot{\bgamma}^R_m (T'_m + |\bgamma^R_m|/c) \cr
& \qquad	+  \left[ |\bgamma^R_m| - \widehat{\bgamma}^R_m \bdot \left( 
	\bm z +  \dot{\bgamma}^R_m (T'_m + |\bgamma^R_m|/c)  \right)  \right] 
	  \bm B_m \bdot \bm v /c  \bigg] /c  \cr
& =  (T'_m - |\bgamma^T_m|/c) \bm B_m \bdot (\bm v - \dot{\bgamma} )/c
	+ \left( R^{RT}_m   - \bm B_m \bdot \left[ \bm p + \bm u T'_m \right] \right)
	\left( 1 + \bm B_m \bdot (\bm v - \dot{\bgamma} )/c \right) /c  \cr
& \qquad- \left( 1 - \widehat{\bgamma}^T_m \bdot \dot{\bgamma}^T_m/c \right) 
	\bigg[ R^{RT}_m - \bm B_m \bdot  (\bm z  + \dot{\bgamma} T'_m )
	+ \widehat{\bgamma}^T_m \bdot \dot{\bgamma}_m (  |\bgamma^T_m|/c)
	- \widehat{\bgamma}^R_m \bdot  \dot{\bgamma}_m ( |\bgamma^R_m|/c) \cr
& \qquad	+  \left[ |\bgamma^R_m| - \widehat{\bgamma}^R_m \bdot \left( 
	\bm z +  \dot{\bgamma}^R_m (T'_m + |\bgamma^R_m|/c)  \right)  \right] 
	  \bm B_m \bdot \bm v /c  \bigg] /c  \cr
& =  T'_m  \bm B_m \bdot (\bm v - \dot{\bgamma} )/c
	- |\bgamma^T_m|  \bm B_m \bdot (\bm v - \dot{\bgamma} )/c^2  \cr
& \qquad
	+ \left( R^{RT}_m   - \bm B_m \bdot \left[ \bm p + \bm u T'_m \right] \right)/c
	+  \left( R^{RT}_m   - \bm B_m \bdot \left[ \bm p + \bm u T'_m \right] \right)
	 \bm B_m \bdot (\bm v - \dot{\bgamma} )/c^2   \cr
& \qquad- 	\bigg[ R^{RT}_m - \bm B_m \bdot  (\bm z  + \dot{\bgamma} T'_m ) \bigg] /c
	+\widehat{\bgamma}^T_m \bdot \dot{\bgamma}^T_m/c 
	\bigg[ R^{RT}_m - \bm B_m \bdot  (\bm z  + \dot{\bgamma} T'_m )	  \bigg] /c  \cr
& \qquad 
	- \left( 1 - \widehat{\bgamma}^T_m \bdot \dot{\bgamma}^T_m/c \right)
	\bigg[ \widehat{\bgamma}^T_m \bdot \dot{\bgamma}_m (  |\bgamma^T_m|/c)
	- \widehat{\bgamma}^R_m \bdot  \dot{\bgamma}_m ( |\bgamma^R_m|/c) \cr
& \qquad	+  \left[ |\bgamma^R_m| - \widehat{\bgamma}^R_m \bdot \left( 
	\bm z +  \dot{\bgamma}^R_m (T'_m + |\bgamma^R_m|/c)  \right)  \right] 
	  \bm B_m \bdot \bm v /c  \bigg] /c  .
\end{align}
where we have written
\begin{equation}
R^{RT}_m =  |\bgamma^T_m| + | \bgamma^R_m| .
\end{equation}
We rearrange \eqref{HRRfocus1} as
\begin{align}	\label{HRRfocus2}
0 &=  	   \bm B_m \bdot \left[(\bm z - \bm p) + ( \bm v -\bm u) T'_m \right] /c  \cr
& \qquad 
	- |\bgamma^T_m|  \bm B_m \bdot (\bm v - \dot{\bgamma} )/c^2
	+  \left( R^{RT}_m   - \bm B_m \bdot \left[ \bm p + \bm u T'_m \right] \right)
	 \bm B_m \bdot (\bm v - \dot{\bgamma} )/c^2   \cr
& \qquad +\widehat{\bgamma}^T_m \bdot \dot{\bgamma}^T_m/c 
	\bigg[ R^{RT}_m - \bm B_m \bdot  (\bm z  + \dot{\bgamma} T'_m )	  \bigg] /c  \cr
& \qquad 
	- \left( 1 - \widehat{\bgamma}^T_m \bdot \dot{\bgamma}^T_m/c \right)
	\bigg[ \widehat{\bgamma}^T_m \bdot \dot{\bgamma}_m (  |\bgamma^T_m|/c)
	- \widehat{\bgamma}^R_m \bdot  \dot{\bgamma}_m ( |\bgamma^R_m|/c) \cr
& \qquad	+  \left[ |\bgamma^R_m| - \widehat{\bgamma}^R_m \bdot \left( 
	\bm z +  \dot{\bgamma}^R_m (T'_m + |\bgamma^R_m|/c)  \right)  \right] 
	  \bm B_m \bdot \bm v /c  \bigg] /c    \cr
&=  	   \bm B_m \bdot \left[(\bm z - \bm p) + ( \bm v -\bm u) T'_m \right] /c  \cr
& \qquad 
	- |\bgamma^T_m|  \bm B_m \bdot (\bm v - \dot{\bgamma} )/c^2
	+  R^{RT}_m    \bm B_m \bdot (\bm v - \dot{\bgamma} )/c^2 
	- \bm B_m \bdot \left[ \bm p + \bm u T'_m \right] 
	 \bm B_m \bdot (\bm v - \dot{\bgamma} )/c^2   \cr
& \qquad +\widehat{\bgamma}^T_m \bdot \dot{\bgamma}^T_m/c 
	\bigg[ R^{RT}_m - \bm B_m \bdot  (\bm z  + \dot{\bgamma} T'_m )	  \bigg] /c  \cr
& \qquad 
	- \left( 1 - \widehat{\bgamma}^T_m \bdot \dot{\bgamma}^T_m/c \right)
	\bigg[ \widehat{\bgamma}^T_m \bdot \dot{\bgamma}_m (  |\bgamma^T_m|/c)
	- \widehat{\bgamma}^R_m \bdot  \dot{\bgamma}_m ( |\bgamma^R_m|/c) \bigg] \cr
& \qquad	- \bigg[ |\bgamma^R_m|   \bm B_m \bdot \bm v /c^2 
- \widehat{\bgamma}^R_m \bdot \left( 
	\bm z +  \dot{\bgamma}^R_m (T'_m + |\bgamma^R_m|/c)  \right)  
	  \bm B_m \bdot \bm v /c^2  \bigg]     \cr
& \qquad  \widehat{\bgamma}^T_m \bdot \dot{\bgamma}^T_m 
	\bigg[ |\bgamma^R_m|   \bm B_m \bdot \bm v /c 
- \widehat{\bgamma}^R_m \bdot \left( 
	\bm z +  \dot{\bgamma}^R_m (T'_m + |\bgamma^R_m|/c)  \right)  
	  \bm B_m \bdot \bm v /c  \bigg] /c^2  .
\end{align}
The first line of \eqref{HRRfocus2} is the condition \eqref{startstopfocus} for
focusing under the start-stop approximation.  
The bottom line of \eqref{HRRfocus2} is of order $(v/c)^2$ and is neglected.  
Thus to leading order in $(v/c)$, the error made in using the start-stop approximation is
\begin{align}	\label{mainerror}
& |\bgamma^T_m|  \bm B_m \bdot  \dot{\bgamma} /c^2
	-  R^{RT}_m    \bm B_m \bdot \dot{\bgamma} /c^2 
	- \bm B_m \bdot \left[ \bm p + \bm u T'_m \right] 
	 \bm B_m \bdot (\bm v - \dot{\bgamma} )/c^2   \cr
& \qquad +\widehat{\bgamma}^T_m \bdot \dot{\bgamma}^T_m/c 
	\bigg[ R^{RT}_m - \bm B_m \bdot  (\bm z  + \dot{\bgamma} T'_m )	  \bigg] /c  \cr
& \qquad 
	- \left( 1 - \widehat{\bgamma}^T_m \bdot \dot{\bgamma}^T_m/c \right)
	\bigg[ \bgamma^T_m \bdot \dot{\bgamma}_m /c
	- \bgamma^R_m \bdot  \dot{\bgamma}_m /c \bigg] \cr
& \qquad	+ \bigg[ 
 \widehat{\bgamma}^R_m \bdot \left( 
	\bm z +  \dot{\bgamma}^R_m (T'_m + |\bgamma^R_m|/c)  \right)  
	  \bm B_m \bdot \bm v /c^2  \bigg]  .
\end{align}
The third line of \eqref{mainerror} is an overall image shift due to the sensor having moved between the time
of transmission and the time of reception.   When this motion is negligible, we have
 $\bgamma^T_m = \bgamma^R_m$ and $\bm B = 2 \hat{\bgamma}$,  in which case
the error \eqref{mainerror} reduces to 
\begin{align}	\label{delayerror}
&=  {1 \over c^2} \bigg( 
	- 4 \left(\hat{\bgamma}_m \bdot \left[ \bm p + \bm u T'_m \right] \right)
	  \hat{\bgamma}_m \bdot (\bm v - \dot{\bgamma} )   
	- 2 \widehat{\bgamma}_m \bdot \dot{\bgamma}_m 
	\bigg[   \hat{\bgamma}_m \bdot  (\bm z  + \dot{\bgamma} T'_m ) \bigg] \cr
& \qquad	+  \widehat{\bgamma}_m \bdot \left( 
	\bm z +  \dot{\bgamma}_m (T'_m + |\bgamma_m|/c)  \right)  
	  2 \hat{\bgamma}_m \bdot \bm v \bigg) .
\end{align}
If, in addition, there is no squint 
($\widehat{\bgamma} \bdot \dot{\bgamma} = 0 =
\bm B \bdot \dot{\bgamma} $), the error \eqref{delayerror} reduces to
\begin{align}	\label{nosquinterror}
\left( {- \bm B_m \bdot \left[ \bm p + \bm u T'_m \right]  + \widehat{\bgamma}_m \bdot \bm z  
	\over c}    \right)    \left( \bm B_m \bdot \bm v /c \right)  
=    \left( {- 2 \hat{\bgamma}_m \bdot \left[ \bm p + \bm u T'_m \right]  + \widehat{\bgamma}_m \bdot \bm z  
	\over c}    \right)    \left( 2 \hat{\bgamma}_m \bdot \bm v /c \right) ,
\end{align} 
which vanishes for stationary targets.  Thus for a stationary scene, we find that misfocusing due to the start-stop approximation can potentially take place only in a squinted system.  Below we compute the magnitude of this effect.  

\subsubsection{Effects on Image}
Resolution in the time delay $\Delta \tau$ is related to range resolution by
$\Delta \tau = 2 \Delta R /c$.   Consequently, to compute the shift in the image due to
the time-delay error \eqref{delayerror}, we multiply \eqref{delayerror} by $c/2$.  


The image is affected only when the error due to the start-stop approximation is greater than a 
resolution cell, because in the backprojection process these contributions will not add correctly. 
 It is well-known \cite{CBbook} that the {\sc sar} cross-range resolution $\delta$ is determined by the aperture $\Delta \theta$ and center frequency $\lambda_0$ as
\begin{equation}	\label{resolution}
\delta = {\lambda_0 \over 2 \Delta \theta} .
\end{equation}
The angular aperture $\Delta \theta$ is determined by the antenna path during which data is collected.   
Most systems are designed so that their down-range and cross-range resolutions are roughly equal.  

We consider the specific example of a low-earth-orbit satellite whose altitude is $300$ km and whose velocity is $8$ km/sec, traversing a straight flight path and forming a spotlight {\sc sar} image of a scene $500$ km from the satellite.    For such a rapid platform speed, the speed of most moving ground targets is insignificant in comparison and is consequently neglected in this example; 
however a similar analysis applies to the case of  very rapidly moving targets. 
We use the assumption $\bgamma^T_m = \bgamma^R_m$. 

\begin{figure}[htbp]
    \centering
    \includegraphics[width=4in]{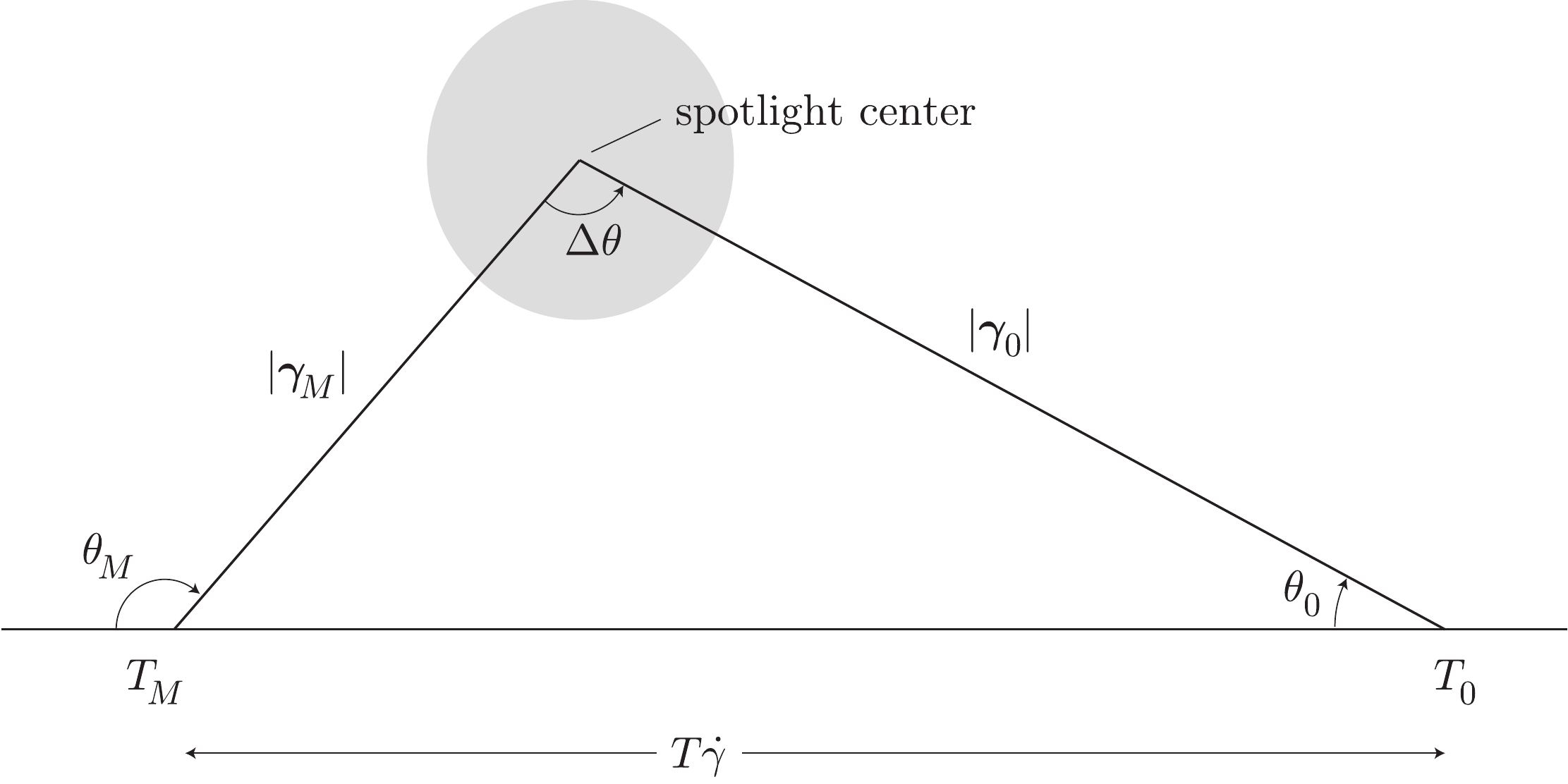} 
    \caption{Spotlight {\sc sar} geometry}
    \label{geometry}
 \end{figure}

We assume data collection starts at pulse $m=0$ and ends at pulse $m=M$.  We denote the angle to the scene center relative to the antenna velocity vector by $\theta_0$ for pulse $m=0$, and  $\theta_M$ when for pulse $m=M$.   See Figure \ref{geometry}.  
 
 The duration of the data collection interval is denoted by $T$, which means for this case that the synthetic aperture is of length $T |\dot{\bgamma}|$, and the system transmits a pulse every $T/(M-1)$ seconds.   To relate this pulse repetition interval, and hence the angular sampling interval, to the system's cross-range resolution, we apply the law of sines to the triangle shown in Fig. \ref{geometry}:  
\begin{equation}	\label{lawofsines}
{\sin \Delta \theta \over T |\dot{\bgamma}| } = {\sin \theta_0 \over |\bgamma_M|} .
\end{equation}
Solving \eqref{lawofsines} for $T$, we find that 
\begin{equation}
T = { |\bgamma_M| \sin \Delta \theta \over |\dot{\bgamma}| \sin \theta_0} 
	\approx { |\bgamma_M| \lambda_0 \over 2 \delta |\dot{\bgamma}| \sin \theta_0} ,
\end{equation}
where we have used the small-$\Delta \theta$ approximation
$ \sin \Delta \theta \approx \Delta \theta$, which from \eqref{resolution} can be written in terms of the system resolution as $ {\lambda_0 / (2 \delta) }$.

\begin{figure}[ht]
\centering    
\includegraphics[width=4in]{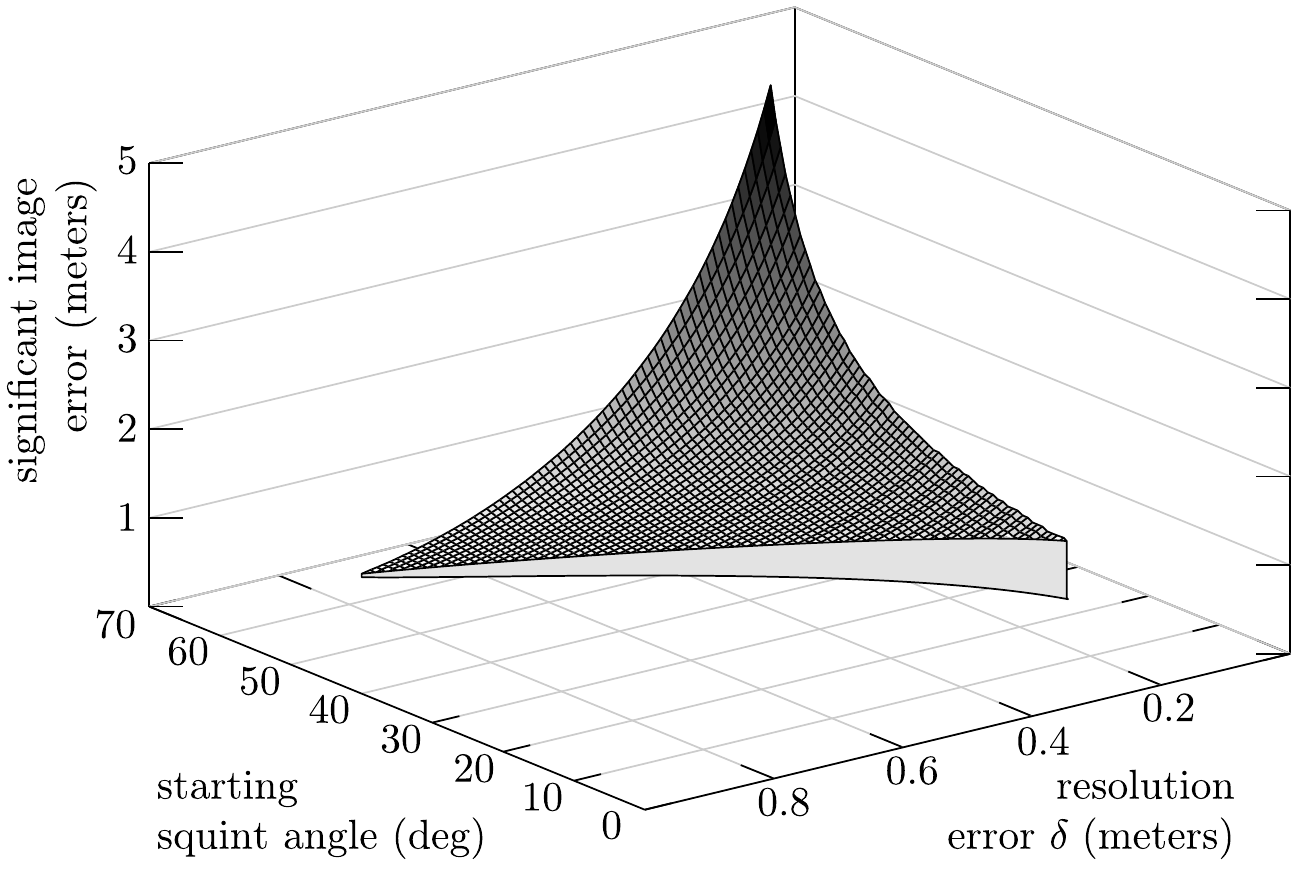} 
\caption{This shows the error as a function of squint and resolution for a target traveling $100$km/hr in the direction opposite to the platform velocity ($8$ km/sec).   We assume $3$-cm wavelength (X-band). }
\label{error}    
\end{figure}

In our simulations, we computed the error \eqref{delayerror} for each pulse, plotted as a function of 
the resolution $\delta$ and the starting angle $\theta_0$.   In Figure \ref{error}, we show only the region in which the error is larger than a resolution cell.

We see from Figure \ref{error} that for a high-resolution system, when the squint angle $\theta_0$ is large, the start-stop approximation could lead to significant image distortion and smearing.  
This can be understood as a consequence of the difference in the way the Doppler scale factor $\alpha$ is treated in the start-stop approximation and the more complete model.  Effectively, in the start-stop approximation, $\alpha$ is set equal to $1$, whereas in the more complete model, we have expanded $\alpha$ in a series of powers of relative velocity.  The error arises from the fact that we correlate an improperly scaled function $f(\alpha t)$ against $f(t)$.  
Contrary to common intuition, which is based on Doppler shifts rather than Doppler scale factors, the mismatch cannot be dismissed as a simple phase shift.  In fact, however, this intuition argues that a very short pulse is sufficiently short that the start-stop approximation is justified, because the duration of its interaction with the target is ``instantaneous".   Such short pulses, however, are necessarily extremely wideband, and Doppler shifts are not appropriate descriptions of the Doppler effect.

\section{Conclusions}
We have developed a theory that can be used to analyze waveform-diversity synthetic-aperture radar.  
This theory has many potential uses; we have shown, for example, that it can be used to analyze errors due to making the commonly-used start-stop approximation.   The results show that even for very short pulses, these errors can be significant in squinted, rapidly-moving, high-resolution imaging systems.   The errors are expected to be worse when longer, more realistic waveforms are used.  

The theory is potentially useful for analysis of imaging artifacts due to moving targets, and 
 for tracking moving targets  \cite{FN}.
 
  It is natural to hope that this theory can provide high-resolution phase-space images of multiple moving targets.  We note, however, that at each $m$, the radar ambiguity function of \eqref{fullpsf2} provides us with two-dimensional information.  The slow-time (position) variable $m$ adds another dimension; thus  we have three-dimensional data.  In the case in which all positions and velocities are restricted to a two-dimensional plane, phase space is four-dimensional.  Thus from three-dimensional data, we are attempting to obtain four-dimensional  information.   This underdetermined imaging situation is currently poorly understood.  Investigations of  similar underdetermined imaging problems were carried out in \cite{WCB,MP}; further exploration is left for the future.

\section{Acknowledgments}

We are grateful to Air Force Office of Scientific Research\footnote{Consequently, the US Government is authorized to reproduce and distribute reprints for governmental purposes notwithstanding any copyright notation thereon. The views and conclusions contained herein are those of the authors and should not be interpreted as necessarily representing the official policies or endorsements, either expressed or implied, of the Air Force Research Laboratory or the US Government.} 
 for supporting this work under the agreements FA9550-06-1-0017 and FA9550-09-1-0013.
 This work was also partially supported by the National Science Foundation under grant CCF-08030672.


%
\end{document}